# Harmonization and interoperability of C-ITS Architectures in Europe and USA

**Areti Kotsi[1], Tom Lusco[2], Evangelos Mitsakis[1], Steve Sill[3]**

[1]Centre for Research and Technology Hellas (CERTH) - Hellenic Institute of Transport (HIT)
[2]Iteris, Inc.
[3]ITS Joint Program Office (JPO) - U.S. Department of Transportation
E-mail: akotsi@certh.gr, ctl@iteris.com, emit@certh.gr, Steve.Sill@dot.gov

## Abstract

Cooperative Intelligent Transportation Systems (C-ITS) constitute technologies which enable vehicles to communicate with each other and/ or with the infrastructure. C-ITS include systems and services which use different components, in order to share and exchange information via diverse communication interfaces. Since various C-ITS deployment initiatives have taken place during the last years all over the world, the necessity to identify a framework, in order to ensure interoperability of C-ITS across borders is currently in sight constituting a challenging task. Such an approach should rely on the deployment of C-ITS services based on common architectures and standards, pursuing harmonization and interoperability. The current paper aims to present the work conducted in the context of a United States (US) - European Commission (EC) collaboration on the development of harmonized and interoperable C-ITS architectures.

***Keywords:*** *Cooperative Intelligent Transportation Systems, architecture, harmonization, interoperability.*

## Περίληψη

Τα Συνεργατικά Ευφυή Συστήματα Μεταφορών (Σ-ΕΣΜ) αποτελούν τεχνολογίες που επιτρέπουν την επικοινωνία μεταξύ οχημάτων και/ ή μεταξύ οχημάτων και υποδομής. Τα Σ-ΕΣΜ περιλαμβάνουν συστήματα και υπηρεσίες που επικοινωνούν, αλληλεπιδρούν και ανταλλάσσουν πληροφορίες μέσω ποικίλων διεπαφών. Οι διάφορες δράσεις και πρωτοβουλίες για την ανάπτυξη των Σ-ΕΣΜ σε παγκόσμιο επίπεδο έχουν οδηγήσει στην ανάγκη για τον καθορισμό ενός πλαισίου το οποίο θα εξασφαλίζει τη διασυνοριακή διαλειτουργικότητα των Σ-ΕΣΜ. Η δημιουργία ενός τέτοιου πλαισίου απαιτεί μια προσέγγιση η οποία θα βασίζεται στη χρήση κοινών αρχιτεκτονικών και προτύπων για την ανάπτυξη των Σ-ΕΣΜ, με σκοπό την εναρμόνιση και τη διαλειτουργικότητα των συστημάτων. Η παρούσα εργασία παρουσιάζει το έργο που πραγματοποιήθηκε για την ανάπτυξη εναρμονισμένων και διαλειτουργικών αρχιτεκτονικών Σ-ΕΣΜ στο πλαίσιο μιας συνεργασίας μεταξύ Ηνωμένων Πολιτειών (ΗΠΑ) και Ευρωπαϊκής Επιτροπής (ΕΕ).

***Λέξεις-κλειδιά:*** *Συνεργατικά Ευφυή Συστήματα Μεταφορών, αρχιτεκτονική, εναρμόνιση, διαλειτουργικότητα.*

## *1. Introduction*

The domain of Intelligent Transportation Systems (ITS) enabling information exchange among vehicles and the roadside infrastructure through digital communication is commonly referred to as Cooperative ITS (C-ITS) (Festag et al., 2014), (Sjoberg et al., 2017). C-ITS successful implementation depends on two factors. Implementation requires the deployment of multiple systems, which are potentially at different development levels, and probably owned and operated by different stakeholders. The end user provided with C-ITS services should be able to use them in a consistent and coherent manner, regardless of the geographic area of delivery and of the stakeholders involved. This level of complexity requires a system architecture and a system engineering approach defining the harmonized combination of all components and expertise for design and development (Jesty & Bossom, 2011).

This paper presents an approach to ensure cross-border C-ITS interoperability by deploying C-ITS services based on common architectures. First an overview on C-ITS architectures in Europe and US is provided, while next sections elaborate on the methodology and the resulting architecture products of this work.

## *2. C-ITS Architectures overview*

### *2.1 European Reference Architectures*

The FRAME Architecture (European ITS Framework Architecture) was created by the EC funded project KAREN (2000). FRAME is not considered so much a model of integrated ITS, as a framework from which specific models of integrated ITS can be created in a systematic and common manner. FRAME is a reference architecture which is broad in scope and not tied to any particular deployment. It is intended to be used as the basis for subsequent architecture definitions; for a project this could include the selection of FRAME functional artifacts, followed by the development of a physical view that allocated that functionality to a physical view specific to the project in question. (Bossom et al., 2000).

The German project CONVERGE (2012 - 2015) created an ITS architecture which was focused on interoperability and economic viability. The CONVERGE architecture is separated in four major structural layers: governance, backend, communication network, and ITS mobile stations (Vogt et al., 2013). MOBiNET (2012 - 2017) was a collaborative project with the objective to simplify the overall process of bringing together mobility service offerings and demand in a common market place. In this view, the architecture and its components create a new ecosystem for drivers, users and providers of transport services (Noyer et al., 2015). The Dutch Integrated Testsite Cooperative Mobility (DITCM) program (2014) had the objective to develop a reference architecture, which could be used as a basis for future ITS deployment projects in the Netherlands (van Sambeek et al, 2015).

The European project Compass4D (2013 - 2015) deployed three services: Energy Efficient Intersection Service, Road Hazard Warning, and Red Light Violation Warning. The reference architecture identifies three main subsystems following the ETSI ITS station architecture (ETSI, E. E. 2010): 1) On-Board Unit (OBU), 2) Roadside Unit (RSU), 3) BO (Traffic Management Center and Pilot Operation Monitoring System) (Alcaraz et al., 2015). The NordicWay project (2015 - 2017) was a pre-deployment pilot of Cooperative Intelligent Transport Systems (C-ITS) in four countries (Finland, Sweden, Norway and Denmark), followed by wide-scale deployment. The NordicWay architecture uses a message queuing approach to transfer messages between the different actors, i.e. Service Providers, Original Equipment Manufacturers (OEMs), and Traffic Message Centers (Sundberg, 2019).

*2.2 US C-ITS Architecture*

ARC-IT provides a common framework for planning, defining, and integrating ITS. ARC-IT, previously called the National ITS Architecture, was first developed in 1996 and has been updated many times over the years to reflect changes in technology and define new ITS services. ARC-IT now includes all content from the Connected Vehicle Reference Implementation Architecture (CVRIA), which until recently was the repository of all C-ITS related architecture material in the United States' reference. ARC-IT includes material content and tools which aim to assist agencies in the development of regional ITS architectures, hence helping regions understand how an individual project fits into a larger regional transportation management context (Iteris, 2017).

ARC-IT is organized around four viewpoints used to describe: Enterprise, Functional, Physical, and Communication. The scope of ARC-IT is defined by a set of ITS Services, meaning transportation services that can be provided through the use of ITS. The range of services covered by ARC-IT is broad and incorporates all the applications of the National ITS Architecture, the CVRIA, and additional ITS services defined internationally. ARC-IT uses the concept of Service Packages to describe the portion of the architecture needed to implement a particular service. Service Packages include the portions of each of the four views needed to describe the service. Service Packages are not intended to be tied to specific technologies, but depend on the current technology and product market in order to actually be implemented. The Areas of Services include Commercial Vehicle Operations, Data Management, Maintenance and Construction, Parking Management, Public Safety, Public Transportation, Support, Sustainable Travel, Traffic Management, Traveler Information, Vehicle Safety, and Weather (Iteris, 2017).

*3. Methodology*

The primary goal of this work is to facilitate application of ARC-IT and its tools to the C-MobILE project. The C-MobILE project (2017 - 2020) focuses on the large-scale demonstration of C-ITS services in eight European cities, Barcelona, Bilbao, Bordeaux, Copenhagen, Newcastle, North Brabant, Thessaloniki, Vigo. Regarding architecture, the main objective was to define a C-ITS reference architecture satisfying stakeholders from public and private parties in an EU context. The first step towards the definition of the architecture was the identification of the use cases (Adali et al., 2018) for the list of C-ITS services deployed within C-MobILE. The use cases were input into the ARC-IT project architecture development process with the goal to produce a representation of the system architecture using ARC-IT language and depictions.

For the purposes of producing the C-MobILE system architecture, the SET-IT tool was used. SET-IT is one of the companion tools for ARC-IT. It is a graphical tool, providing the user with visual feedback and tools necessary to manipulate service package physical and enterprise diagrams, develop communications stack templates, specify standards at all protocol layers, and export that information in a variety of forms and formats. The tool was considered appropriate as it is project-focused and can apply to an individual project deployment. The whole process produced artifacts which served as pre-design documentation for the C-MobILE system architecture, enabling the identification of functionalities within subsystems and interfaces between subsystems.

## *4. Architecture products*

The representation of the C-MobILE system architecture, produced by SET-IT, is comprised of several Physical View diagrams depicting the Physical Objects (p-objects), i.e. persons, places, or things participating in C-ITS, as well as the information flows (highest-level definition of interfaces) between them. Each diagram describes the interactions between the support, center, field, traveler and vehicle systems involved in each C-ITS service. P-objects are defined as follows:

- Center (green rectangle): Element providing application, management, administrative, and support functions from a fixed location not in proximity to the road network.
- Field (orange rectangle): Infrastructure proximate to the transportation network performing surveillance, traffic control, information provision and local transaction functions.
- Support (light green rectangle): Center providing a non-transportation specific service.
- Traveler (yellow rectangle): Equipment used by travelers to access transportation services pre-trip and en-route.
- Vehicle (blue rectangle): Vehicles, including driver information and safety systems applicable to all vehicle types.

An information flow contains one or more dialogs that contain one or more messages (which may be unicast, multicast or broadcast); each message contains one or more data elements. Arrows indicate the direction data flows, though messages may go in both directions as part of request-response type interactions. Line colors indicate whether the key messages are expected to be authenticable (green) or authenticable and obfuscated (red) or neither (black). Alphanumeric codes indicate time and spatial relevance (A-D, 1-4 respectively). Information flow initiators are indicated by the presence of a small white square affixed to the flow, located inside the initiating p-object. A small bar perpendicular to the flow line on the side of the line opposite the arrow indicates that the flow is expected to includes some form of acknowledgement.

The following Physical View diagrams represent the system architecture of the C-MobILE services including the whole number of use cases which address the needs of three stakeholders' categories: 1) Customers/ End-users, 2) Technology Providers, and 3) Legal Authorities.

Rest Time Management supports managing the working hours of drivers engaged in the transport of goods and passengers by road. The use case represents the display of available parking spots along the route of a truck (or commercial vehicle) driver, at a certain frequency.

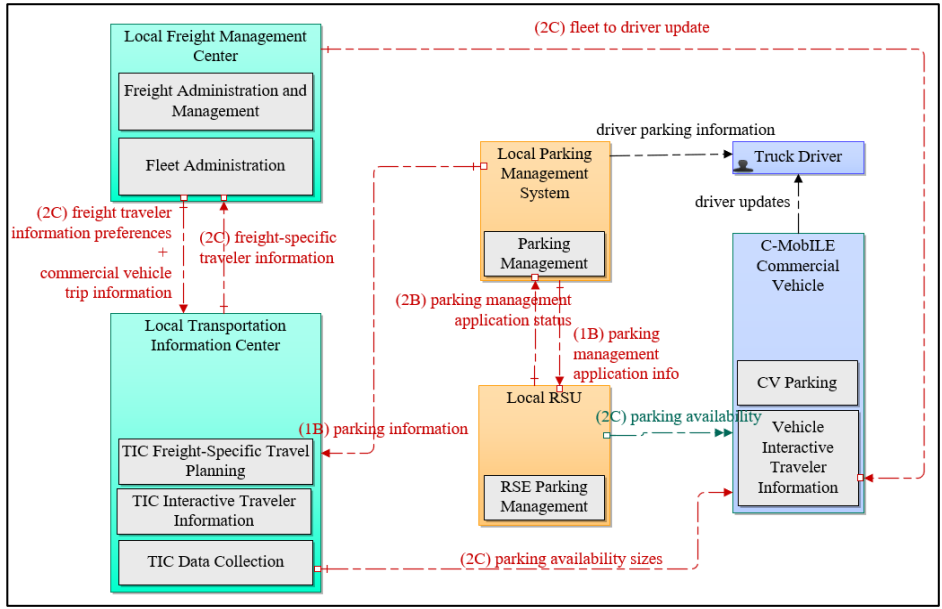

*Figure 1:* Rest Time Management

Motorway Parking Availability provides motorway parking availability information and guidance for truck drivers to make informed choices about available parking places. The use cases represent the provision of information on parking lots location, availability and services (via internet and via I2V), on reservations/ truck parking spaces released by users, and on guiding the truck in a port.

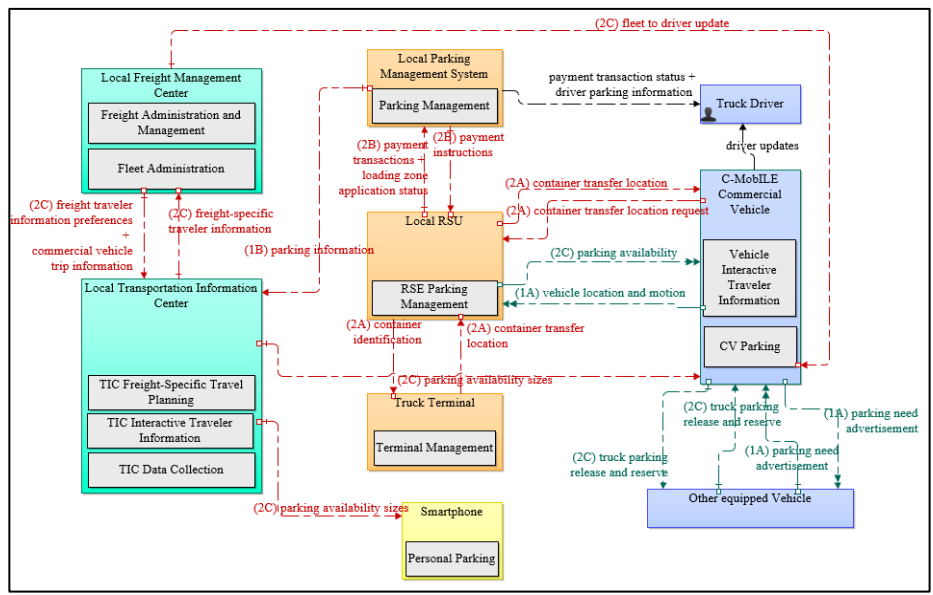

*Figure 2:* Motorway Parking Availability

Urban Parking Availability provides parking availability information and guidance for drivers to make informed choices about available parking places. The use cases represent the provision of information about reservations/ vehicles' parking space released by users, about on-street parking availability for both urban freight and private cars.

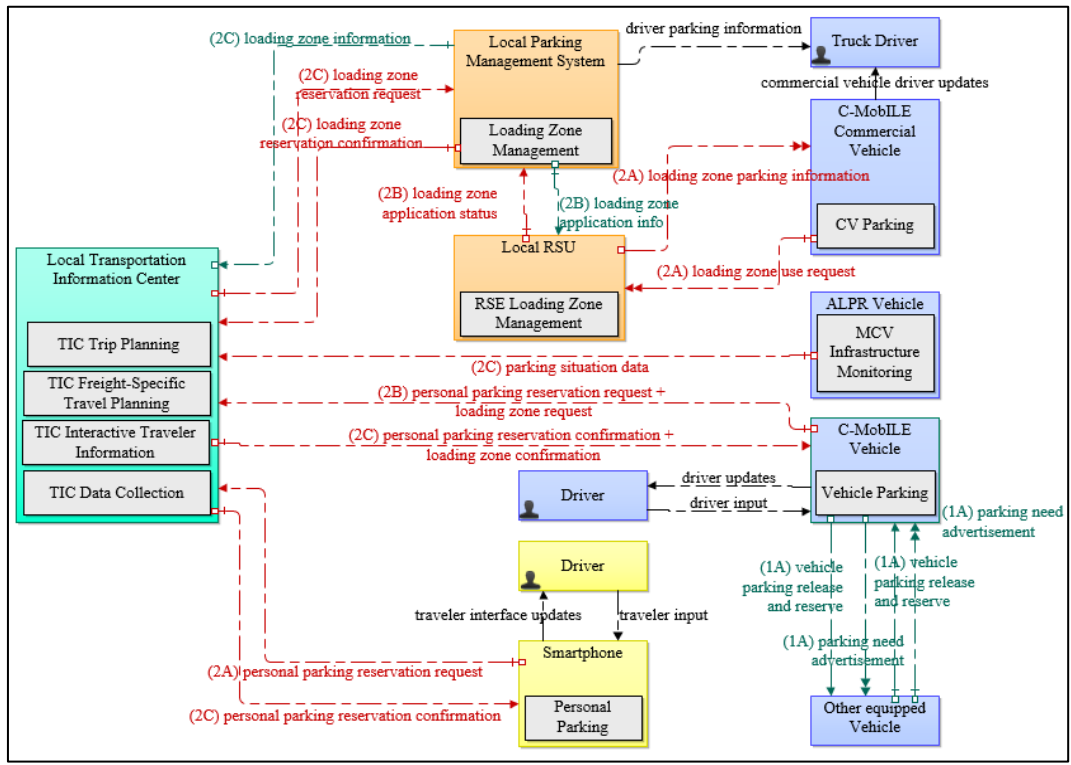

*Figure 3: Urban Parking Availability*

Road Works Warning provides in-vehicle information and warnings about road works, changes to the road layout and applicable driving regulations. The use cases represent the provision of such information via cellular communications (2G/ 3G/ 4G) and IEEE 802.11p.

*Figure 4:* Roads Works Warning

Road Hazard Warning informs drivers in a timely manner of upcoming, and possibly dangerous events and locations. The use cases represent the provision of hazardous location notification, traffic and weather conditions warning.

*Figure 5:* Road Hazard Warning

Emergency Vehicle Warning provides in-vehicle information and warnings about approaching emergency vehicles. The use cases represent the provision of such information via cellular communications (2G/ 3G/ 4G) and via IEEE 802.11p.

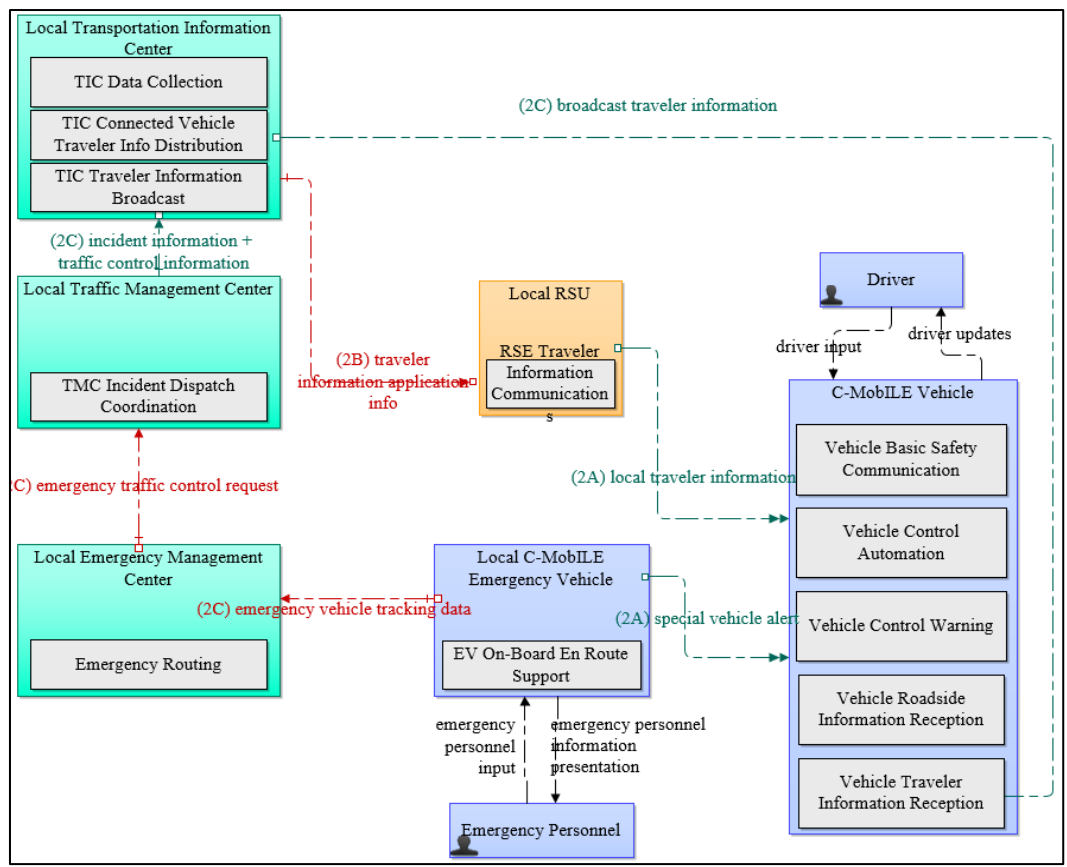

*Figure 6:* Emergency Vehicle Warning

Warning System for Pedestrian aims to detect risky situations involving pedestrians, allowing the possibility to warn vehicle drivers. The use case addresses safe travelling experience by warning signage.

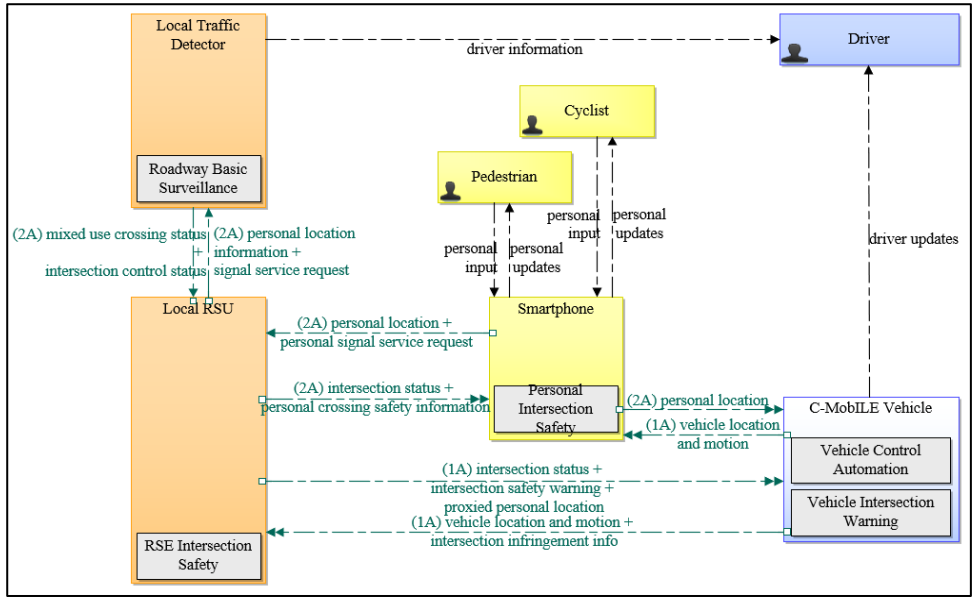

*Figure 7:* Warning System for Pedestrians

Green Priority aims to change the traffic signals status along the route of a priority vehicle, halting conflicting traffic. The use case address green priority for designated vehicles.

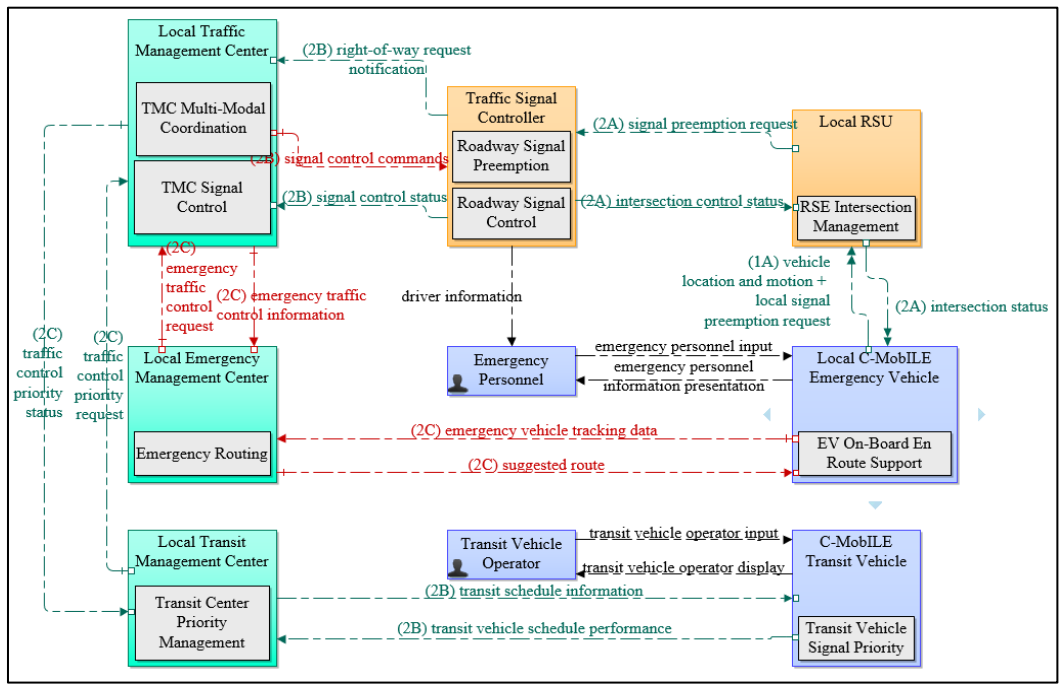

*Figure 8:* Green Priority

GLOSA provides vehicle drivers an optimal speed advice when they approach a controlled intersection equipped with traffic lights. The use case represents optimized driving experience with GLOSA.

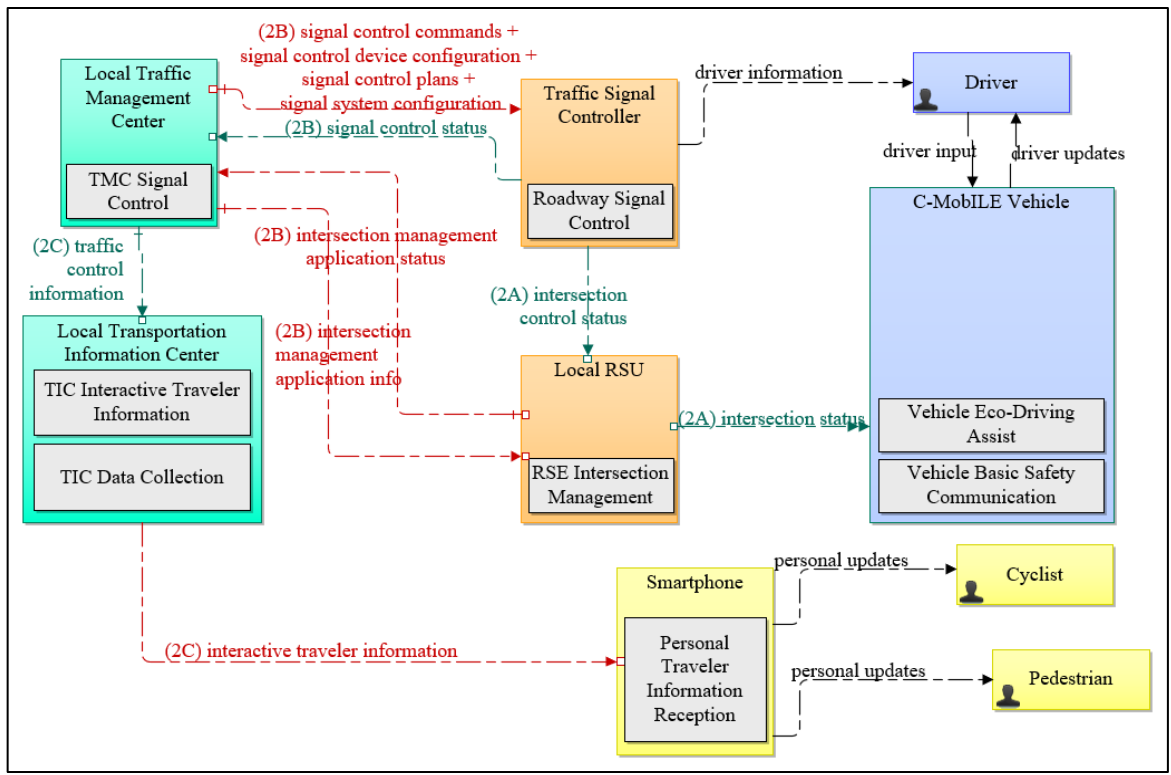

*Figure 9: GLOSA*

Cooperative Traffic Light for VRUs addresses traffic signal timing and priority assignment based on detection of VRUs using fixed sensors and portable devices. The use cases represent traffic light prioritization for designated VRUs and cooperative traffic light with VRU counting.

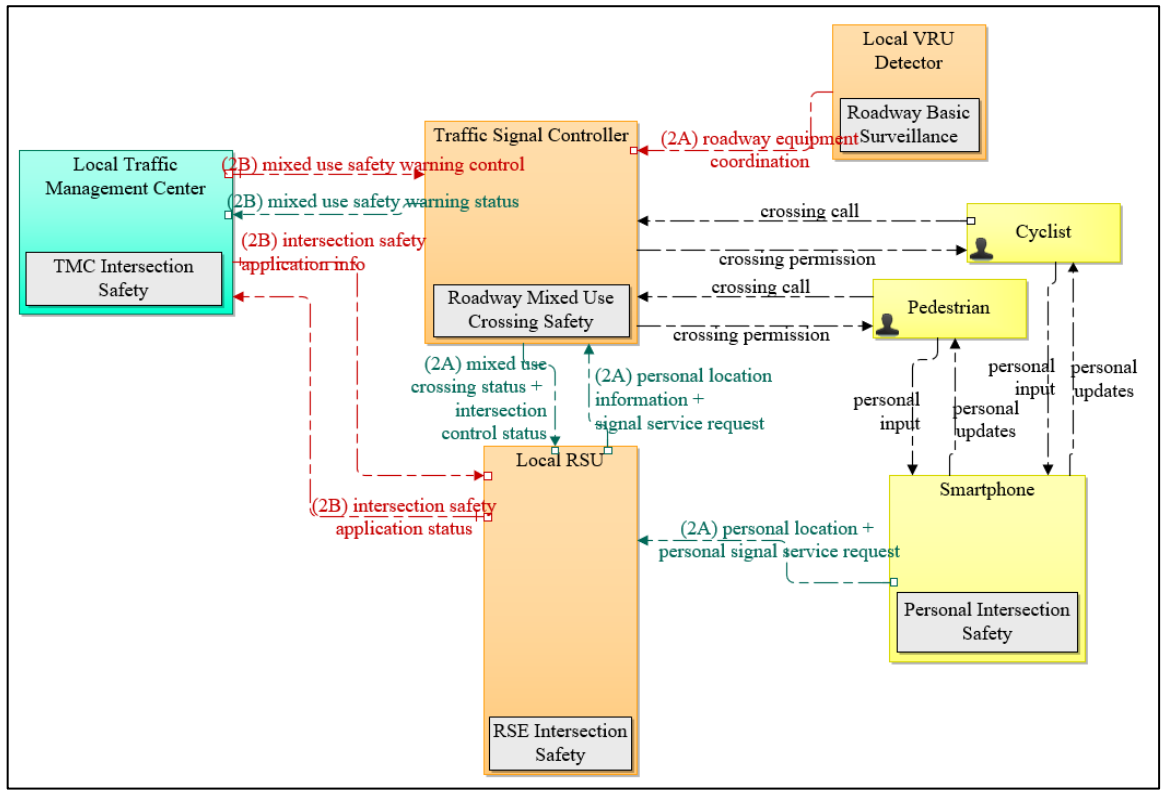

*Figure 10:* Cooperative Traffic Light for VRUs

Flexible Infrastructure aims to interchange information about the lanes provided to the traffic users according to the time of the day. The use cases represent dynamic lane management, lane status information, and reserved lane (with/ without the use of probe vehicle data).

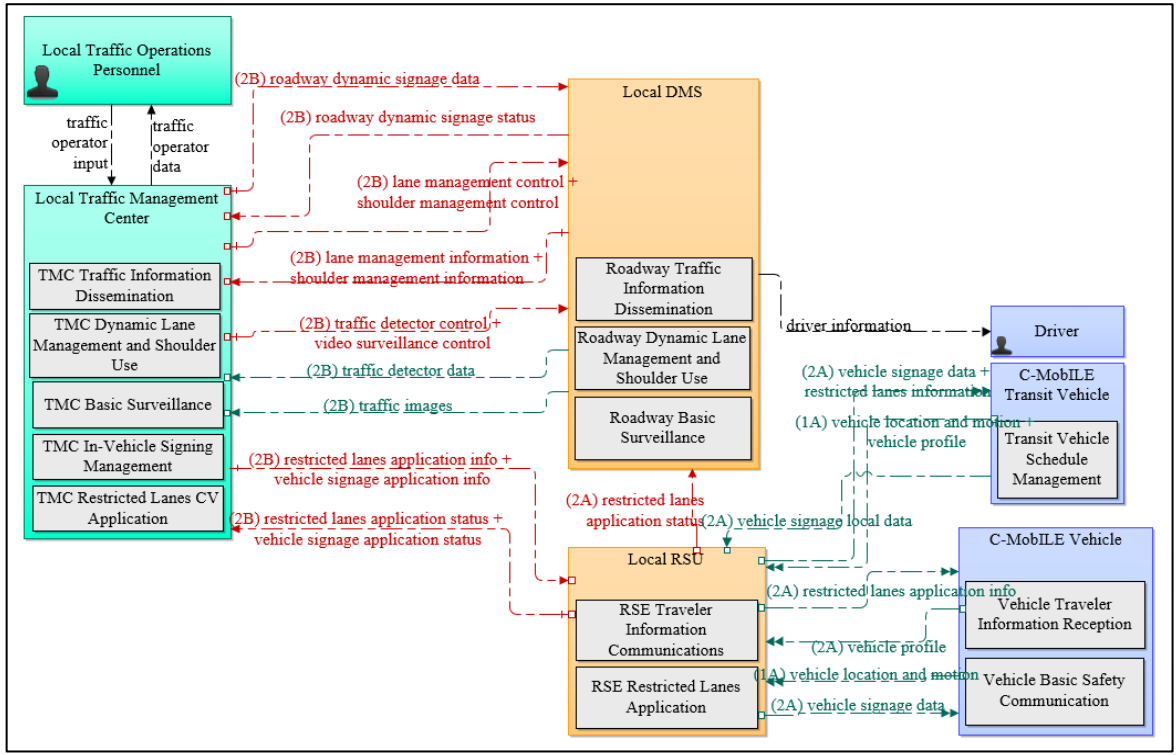

*Figure 11: Flexible Infrastructure*

In Vehicle Signage shows both static and dynamic information of road signs inside the vehicle. The use cases represent the provision of information on dynamic/ static traffic signs via cellular communication (2G/ 3G/ 3G) and IEEE 802.11p.

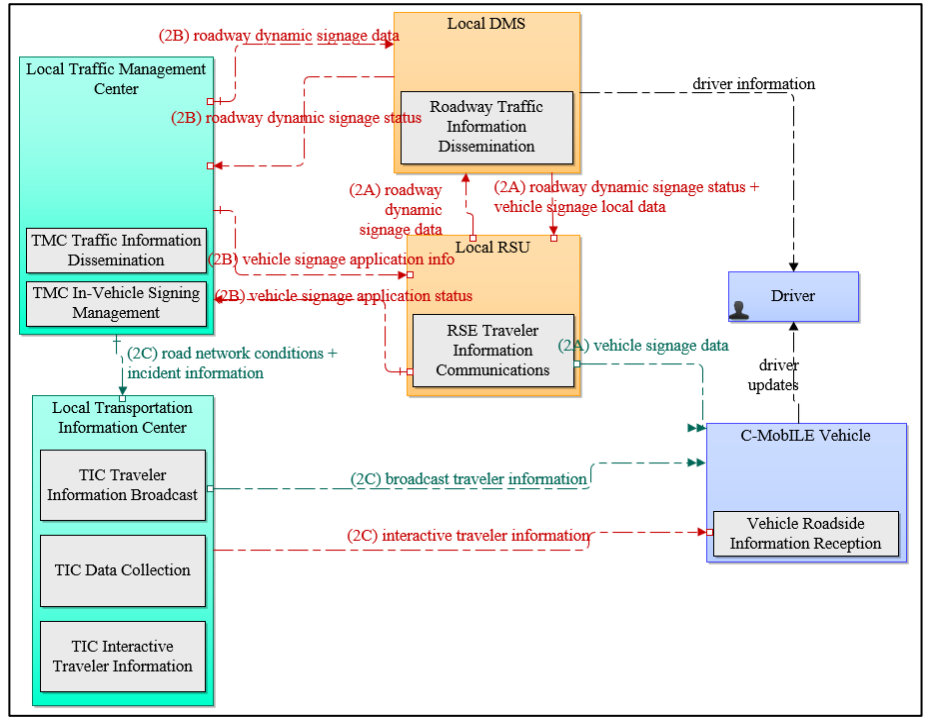

*Figure 12: In-Vehicle Signage*

Mode & Trip Time Advice aims to provide travelers with an itinerary for a multimodal passenger transport journey, taking into account real-time and/ or static multimodal journey information. The use cases represent provision of information for event visitors, drivers, and cyclists.

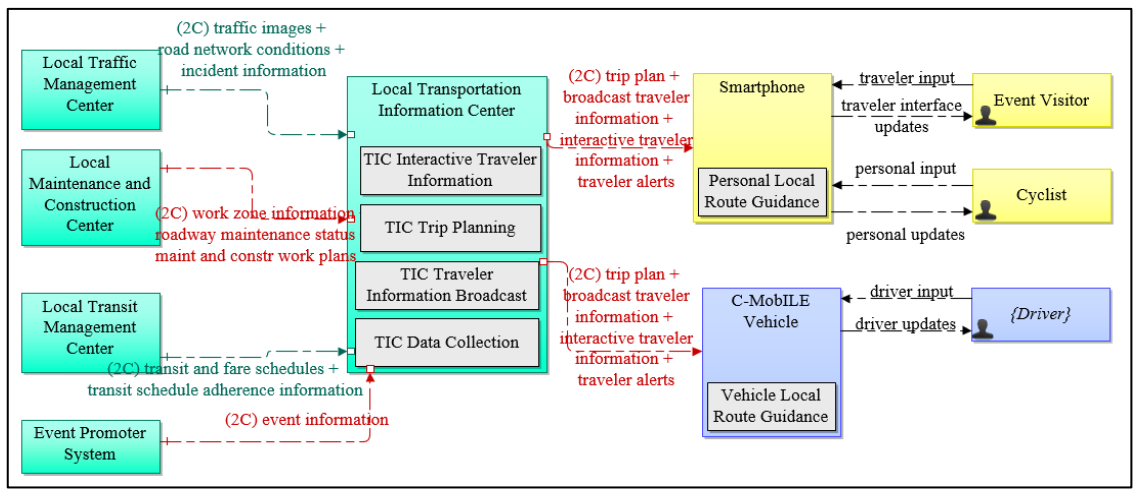

*Figure 13: Mode & Trip Time Advice*

Probe Vehicle Data is data generated by vehicles. The use cases include basic and extended probe vehicle data.

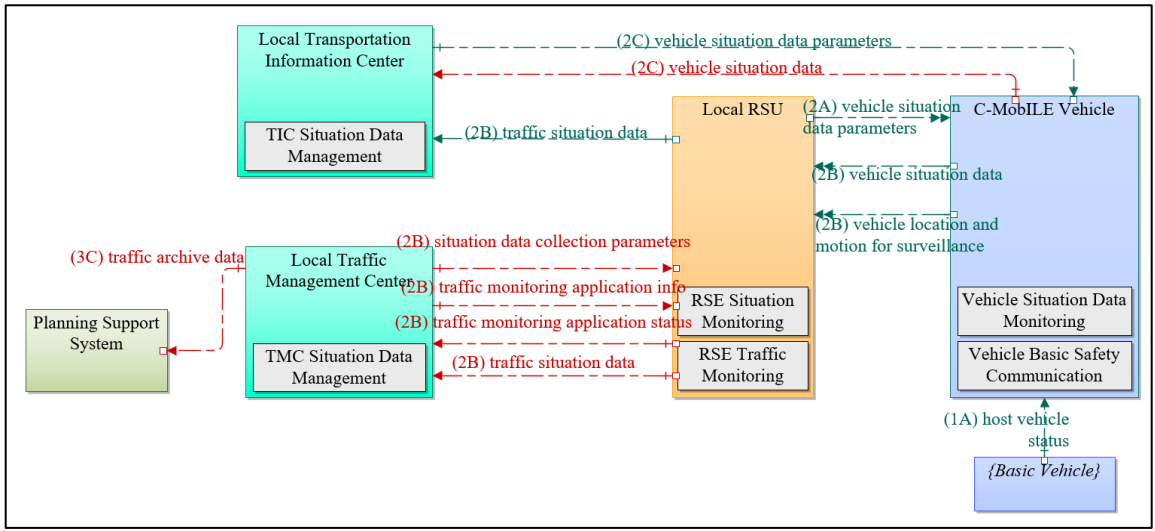

*Figure 14: Probe Vehicle Data*

Emergency Brake Light aims to avoid rear end collisions, which can occur if a vehicle ahead suddenly brakes, especially in dense driving situations or in situations with decreased visibility. The use case addresses emergency electronic brake lights.

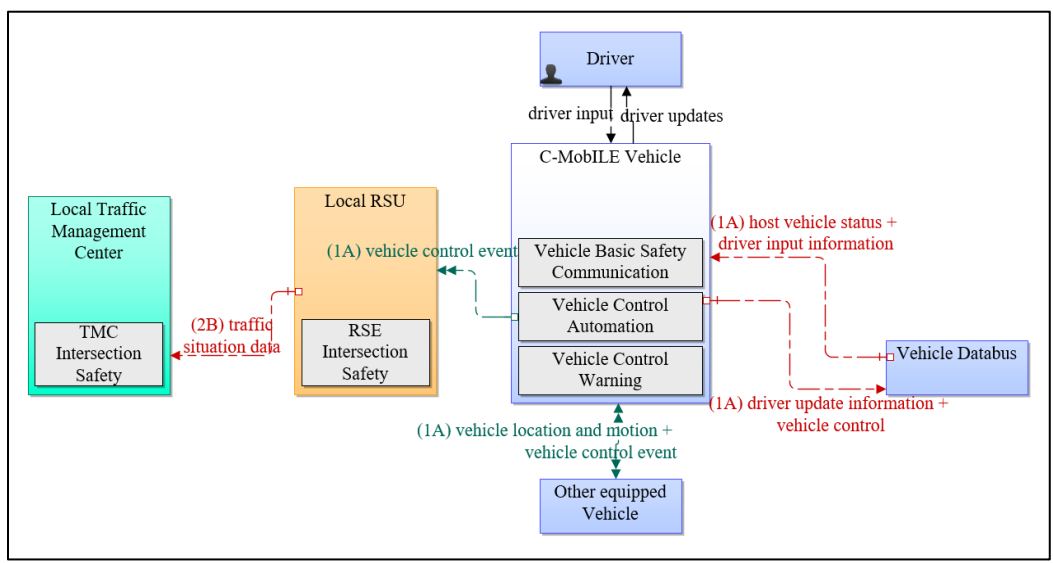

*Figure 15: Emergency Brake Light*

Cooperative Adaptive Cruise Control (CACC) ensures smooth driving of vehicles with enabled CACC function or platooning for driving through C-ITS equipped intersections. The use cases include CACC passenger vehicles approaching urban/ semi-urban environment.

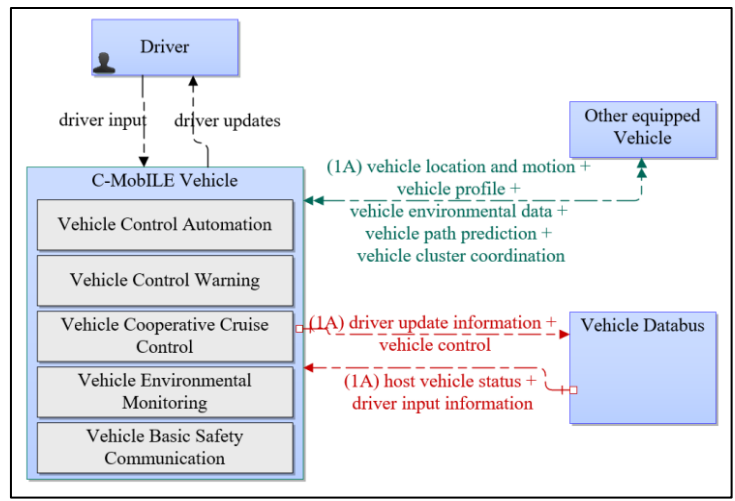

*Figure 16: CACC*

Motorcycle approaching indication warns drivers that a motorcycle is approaching/ passing. The use cases include V2V and V2I two-wheeler approaching warning.

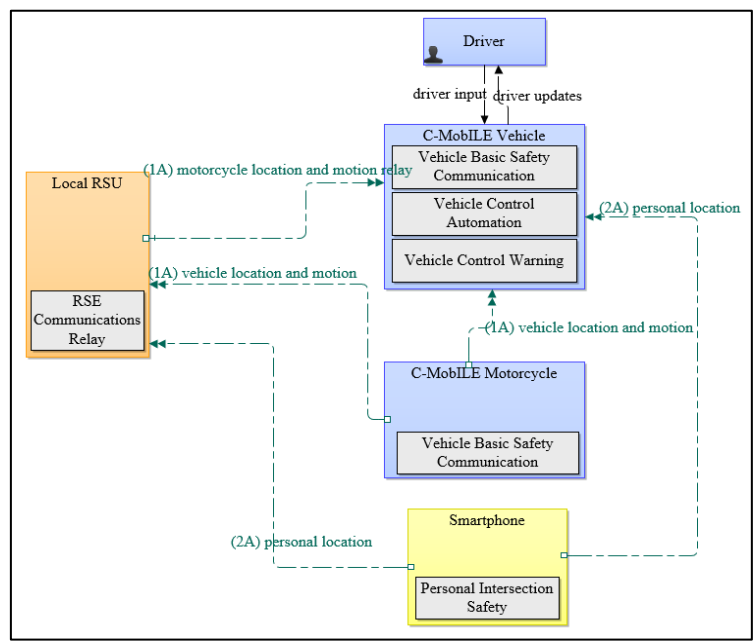

*Figure 17: Motorcycle Approaching Indication*

Slow or Stationary Vehicle Warning provides timely in-car driving assistance information on a slow/ stationary vehicle(s) downstream of the current position and in the driving direction of the vehicle.

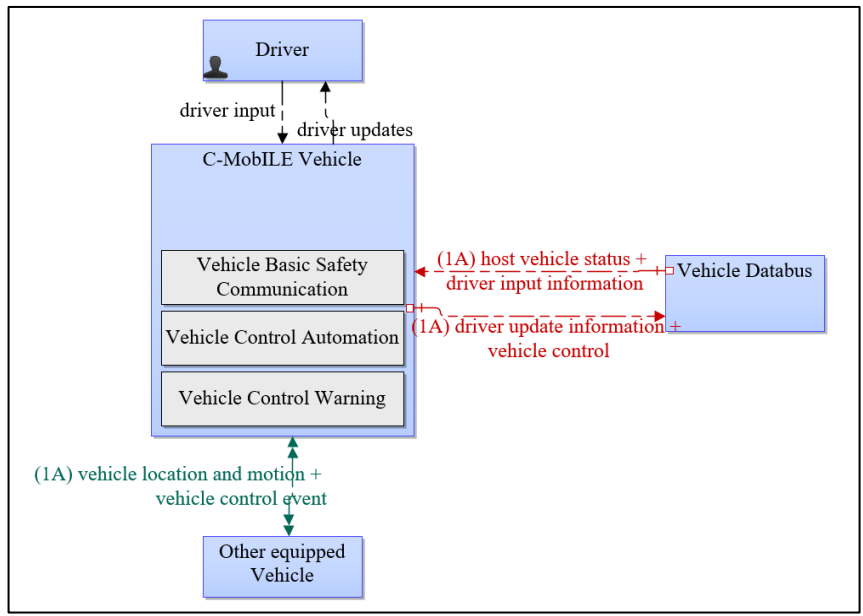

*Figure 18: Slow or Stationary Vehicle Warning*

Blind spot detection / warning aims to detect and warn drivers about other vehicles of any type located out of sight. The use case includes digital road safety mirror addressing also VRUs.

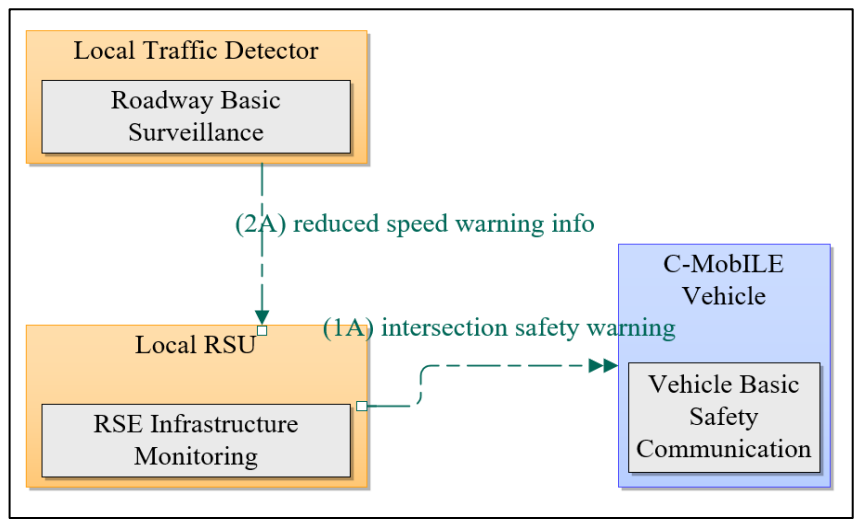

*Figure 19: Blind Spot Detection/ Warning*

## 5. Conclusions

Progress in the innovative C-ITS domain depends strongly upon the active engagement of a vast number of stakeholders, such as communication and mobility services providers, road operators, traffic managers, fleet managers, and vehicle manufacturers. For a successful cooperation the establishment of a framework covering systems definition, and reliable and secure information exchange is necessary. The objective of this work is to address this challenge through interoperable C-ITS architectures. In this context a US - EC collaboration on C-ITS architectures is presented. The outcome of this work includes the representation of the C-MobILE system architecture by facilitating the application of ARC-IT and the SET-IT tool. This cooperation on common C-ITS system architecture development promotes the establishment of a C-ITS framework with the potential to introduce an overall C-ITS architecture ensuring interoperability, and maximizing benefits along the whole value chain of stakeholders.

## 6. References-Bibliography